# A Local Deterministic Model of Quantum Spin Measurement

by

T.N. Palmer

ECMWF
Shinfield Park
RG2 9AX
UK

# Abstract

The conventional view, that Einstein was wrong to believe that quantum physics is local and deterministic, is challenged. A parametrized model, "$Q$", for the state vector evolution of spin-1/2 particles during measurement is developed. $Q$ draws on recent work on so-called "riddled basins" in dynamical systems theory, and is local, deterministic, nonlinear and time asymmetric. Moreover, the evolution of the state vector to one of two chaotic attractors (taken to represent observed spin states) is effectively uncomputable. Motivation for considering this model arises from speculations about the (time asymmetric and uncomputable) nature of quantum gravity, and the (nonlinear) role of gravity in quantum state vector reduction.

Although the evolution of $Q$'s state vector cannot be determined by a numerical algorithm, the probability that initial states in some given region of phase space will evolve to one of these attractors, is itself computable. These probabilities can be made to correspond to observed quantum spin probabilities. In an ensemble sense, the evolution of the state vector to an attractor can be described in by a diffusive random walk process, suggesting that deterministic dynamics may underlie recent attempts to model state vector evolution by stochastic equations.

Bell's theorem and a version of the Bell-Kochen-Specker quantum entanglement paradox, as illustrated by Penrose's "magic dodecahedra", are discussed using $Q$ as a model of quantum spin measurement. It is shown that in both cases, proving an inconsistency with locality demands the existence of definite truth values to certain counterfactual propositions. In $Q$, these deterministic propositions are physically uncomputable, and no non-algorithmic mathematical solution is either known or suspected. Adapting the mathematical formalist approach, the non-existence of definite truth values to such counterfactual propositions is posited. No inconsistency with experiment is found. As a result, it is claimed that $Q$ is not constrained by Bell's inequality, locality and determinism notwithstanding.

# 1. Introduction

Einstein believed that physical theory should be both deterministic and local. His objections: "God does not play dice", and, "quantum theory cannot be reconciled with the idea that physics should represent a reality in time and space, free from the spooky action at a distance" are two of the most famous remarks about 20th century physics.

However, the possibility of describing state vector evolution during measurement as deterministic does not appear reconcilable with the seemingly stochastic nature of macroscopic observations on individual quantum states. In recent years, stochastic models have been developed which describe the evolution of a quantum state during the process of measurement (eg Diosi, 1989; Gisin, 1984; Ghirardi et al, 1986; Pearle, 1993; Percival, 1994). These equations describe, for example, how the interaction of a quantum system with its environment leads to an irreversible diffusion towards an eigenstate of the measurement operator.

Moreover, Bell's (1964) celebrated inequalities also appear to rule out a local deterministic quantum theory, and it is now a "textbook" result (eg Rae, 1992) that quantum theory is inherently nonlocal. However, for proponents of the belief that physics should have an objective reality, nonlocality raises profound conceptual problems, specifically the difficulty in reconciling quantum reality with Lorentz invariance (a measurement "here" instantaneously determining a quantum state "there"). For example, Penrose (1987) remarks that whilst one could simply abandon relativistic invariance in formulating a picture of quantum reality, which would not lead to any conflict with experiment, "there is conflict with the spirit of relativity". To abandon a relativistic view of our picture of reality, Penrose argues, would be to ignore the insight that Einstein impressed on us in 1905. However, despite these difficulties, Penrose (1994a) concludes that "nonlocality seems to be here to stay".

Notwithstanding these remarks, the purpose of this paper is to challenge the conventional view that Einstein was wrong to insist that physical theory be local and deterministic. Motivated by the possible role of gravity on quantum state vector reduction, and by recent work in dynamical systems, an idealised ("parametrized") local deterministic model, "$Q$", of quantum spin-1/2 measurement is proposed. The phase-space of $Q$ has two symmetric (chaotic) attractor sets whose basins cover phase space, but are intermingled, in the sense that neither basin has any open sets. Whilst $Q$'s dynamics are deterministic, in the sense that there are no stochastic elements, this property of intermingling makes $Q$ extremely unpredictable. Indeed, the phase-space evolution of the state vector towards either of the attractors, which are taken to correspond to observable states, can be made to be effectively uncomputable. In an ensemble sense it can be shown that state-vector evolution mimics stochastic diffusion precisely, and the probability of evolution towards one attractor or the other can be made to be consistent with quantum spin-1/2 statistics.

A critical factor in the application of $Q$ to the physical world concerns the question of existence of exact integrations of the governing dynamical equations, which are required to determine which basin of attraction a given initial phase-space point lies, and are postulated to be physically associated with the process of measurement. By construction, such exact integrations cannot be obtained from physically-based algorithms. Neither do they appear to be obtainable from other non-algorithmic mathematical approaches. Hence, from this perspective, their existence is ambiguous. However, from a physical point of view, if an exact integration corresponds to a possible physical measurement, its value clearly must exist. Nevertheless, we show that in the derivation of Bell's theorem and a second quantum entanglement theorem, there are uncomputable propositions which do not correspond to any possible physical measurement. We define an approach, "Natural Formalism" (NF), to this problem of existence, which could be thought of as an adaptation of Mathematical Formalism to physical theory. In NF, unprovable and untestable propositions are denied definite truth values. Such an approach denies the validity of counterfactual statements (Stapp, 1994). As such, in an NF interpretation, $Q$ is not constrained to satisfy Bell's inequalities. Hence, whilst it is not claimed that $Q$ is complete, we conclude that in a deterministic model, the assumption of locality does not necessarily lead to inconsistency.

A detailed account of the primary motivation for the proposed model is given in section 2, and the model itself is outlined in section 3. The NF approach proposed in this paper is discussed in section 4, and provides some reconciliation between the Bohr and Einstein positions on quantum reality, but is antithetical to the many-worlds interpretation. The application to Bell's theorem and to Penrose's "magic dodecahedra" are given in sections 5 and 6 respectively. Concluding remarks are made in section 7.

## 2. Some preliminaries

In this section we outline the principal motivations for considering the model described in section 3 to parametrize quantum measurement.

### (a) Gravity and topological undecidability

The concept of computability in physical theory has been discussed by Geroch and Hartle (1986). They argue that conventional theories of physics have the property that all measurable quantities specified by these theories are computable, ie these quantities can be calculated to arbitrary accuracy, using algorithmic approximations to the governing equations.

On the other hand, such computability may break down when quantum gravitational effects are included in fundamental physical theory. In particular Geroch and Hartle (1986) consider an observable $A$, in a theory of quantum gravity for closed cosmologies, as a functional of geometries $G$ on compact 4-manifolds. They write the expectation value $<A>$ using the "sum over histories" formulation, where the sums are over all compact 4-geometries, including all possible topologies for the 4-manifolds.

Consider an algorithmic approach to finding $<A>$. Let $G_n$ be an $n$-vertex simplicial approximation to $G$. The expectation value $<A_n>$ evaluated by summing over such $G_n$ could be expected to give an arbitrarily good approximation to $<A>$ for sufficiently large $n$. However, it is not possible to decide algorithmically whether two simplicial 4-manifolds are homeomorphic. This problem is isomorphic to the halting problem for finite strings (Haken, 1973). Hence, in evaluating the sum over histories, it is impossible to eliminate all duplications; cases in which there is more than one simplicial representation of a single 4-manifold with given topology. Hence, in general, a sum over functionals of all $n$-vertex simplicial manifolds cannot be made to correspond to the sum over all distinct 4 manifolds.

*(b) Gravity and irreversibility*

The notion that at the quantum level, gravitation must have some explicitly time-asymmetric component, is supported by studies of classical space-time singularities. In particular, the curvature singularities associated with the collapse of matter are dominated by the Weyl component of the curvature tensor. By contrast, the curvature singularities associated with symmetric Robertson-Walker cosmologies are dominated by the (tidal-free) Ricci component. This has lead Penrose (1979) to speculate that the characteristics of initial (big-bang) and final (big-crunch) cosmological singularities are of fundamentally different character, the former having Ricci-dominated curvature, the latter having Weyl-dominated curvature. This "Weyl Curvature Hypothesis" leads directly to a preferred direction for time, and to a "second law of thermodynamics", since Weyl-dominated singularities must be associated with highly "clumped" states of matter, and therefore, from a gravitational point of view, with high entropy. By contrast, Ricci-dominated singularities are associated with strongly symmetric states, and therefore with low entropy.

In classical theory, singularities are excluded from the space-time manifold, and this arrow of time arises from asymmetries in manifold boundary conditions rather than from irreversibility in the equations of motion. As such, the time symmetry of classical general relativity is not inconsistent with the Weyl Curvature Hypothesis. On the other hand, a quantum theory of gravitation, which must "internally" account for the classical curvature singularities, must necessarily be time asymmetric, or irreversible, to be consistent with the Weyl Curvature Hypothesis. The possible time-asymmetric nature of quantum gravity is consistent with its possible role in the irreversible process of quantum state reduction.

*(c) Gravity, nonlinearity and quantum state vector reduction*

A number of authors have argued that quantum gravitation effects may be fundamental in understanding the quantum measurement problem ( eg Károlyházy et al, 1986; Penrose,

1987, Diosi, 1989, Ghirardi et al, 1990).

Recently, Penrose (1994b) based on earlier ideas of Diosi (1989) has suggested that state reduction might be viewed as a gravitationally-induced instability between states which might otherwise exist in linear superposition. A timescale associated with this instability would depend on the energy difference between these states.

A very approximate calculation of this effect is given by Penrose (1994b) who considers two 'lumps' of matter of mass $m$ and radius $a$, and considers the energy it would take to displace one lump from the other, considering only gravitational effects. This gives a timescale $a/m^2$ ( in geometric units) which Penrose equates with a quantum state reduction time. For an elementary particle this timescale is over 10 million years, for a microscopic water droplet the timescale is fractions of a second. A similar scaling has been found by Percival (1994) in his model of gravitationally-induced primary state diffusion.

This is an area where further work is needed to give more definitive results. In this paper, we assume that the possibility of gravitationally-induced state-vector reduction is plausible. In providing support for Einstein's belief in locality and determinism, it seems somehow appropriate that gravitation may be the crucial ingredient for a relativistically-invariant picture of quantum reality.

*(d) Uncomputability, irreversibility and riddled attractor basins in dynamical systems theory.*

The unpredictable nature of dissipative chaotic dynamical systems is familiar, following the work of the meteorologist Lorenz (1963). An approach to the formulation of deterministic irreversible dynamical systems with essentially undecidable characteristics has been developed by Alexander et al (1992) and Ott et al (1993, 1994) who consider a class of dissipative dynamical systems containing a chaotic attractor set $C$ for which all points in the basin of attraction $b(C)$ of $C$ have pieces of another attractor basin arbitrarily near. That is, if $r \in b(C)$, then for every $\varepsilon > 0$ there are displacements $\delta$ where $|\delta| < \varepsilon$, such that the point $r + \delta$ is in the basin of attraction of another attractor, and the set of these points has positive Lesbesgue measure. Such basins are called riddled. Alexander et al (1992) conclude that riddled basins are not exceptional for certain classes of dynamical system.

According to Ott et al (1994), conditions under which riddled basins occur for a $d$-dimensional dynamical system are as follows:

(i) There is an invariant subspace $M$ (whose dimension is less than $d$).
(ii) There is a chaotic attractor $C$ for initial conditions in $M$.
(iii) The Lyapunov exponents associated with infinitesimal perturbations to typical orbits on $C$, in the directions transverse to $M$, are negative.
(iv) At least one of these transverse Lyapunov exponents experiences positive finite-time fluctuations.
(v) There is another attractor not in $M$.

A specific example of such a dynamical system, based on a simple pair of ordinary differential equations, has been recently found by Sommerer and Ott (1993) who consider the position vector $r=(x,y)$ of a unit-mass particle moving in the $x$–$y$ plane. The particle acceleration is given by the sum of a frictional force, the gradient of a potential function and a periodic external force in the $x$-direction. The particle's equation of motion is

$$\frac{d^2x}{dt^2} = -\gamma \frac{dx}{dt} - \nabla_x V(x,y) + p \sin \omega t \qquad \frac{d^2y}{dt^2} = -\gamma \frac{dy}{dt} - \nabla_y V(x,y) \qquad (2.1)$$

where the potential is

$$V(x,y) = V_{SO}(x,y) = (1-x^2)^2 + (x+\bar{x})y^2. \qquad (2.2)$$

In phase space, these equations have an invariant manifold at $y=0$, which, for Sommerer and Orr's parameter values $\gamma=0.05, p=2.3, \omega=3.5$ contains the Duffing twin-well (chaotic) attractor (Thompson and Stewart, 1991). This attractor can be generated from the equation

$$\frac{d^2x}{dt^2} = -\gamma\frac{dx}{dt} - \nabla V(x) + p\sin\omega t \qquad (2.3)$$

where

$$V(x) = V_D(x) = (1-x^2)^2 \qquad (2.4)$$

and is illustrated in the phase space ($x$, $dx/dt$) in Fig 1. Associated with (2.1), (2.2), there is a second attractor at $y=\infty$.

Sommerer and Ott (1993) demonstrate numerically that, for suitable values of $\bar{x}$, the basin of attraction of the Duffing attractor in the $y=0$ invariant manifold is riddled with respect to the second attractor at infinity. (We illustrate numerically this behaviour on a related equation set defined below in this section.)

The value of $\bar{x}$ in (2.2) is important. When $x > -\bar{x}$, the potential gradient in the $y$ direction forces the state vector to the $y=0$ invariant plane; when $x < -\bar{x}$, the potential gradient forces the state vector towards $y=\infty$. For large $\bar{x}$, the potential almost always forces the state vector to $y=0$. When $\bar{x} < \bar{x}_{cr} = 1.7887...$ (for parameter values given above), a small $y$-perturbation to a typical trajectory in the $y=0$ invariant manifold, will cause the state vector to experience the repelling behaviour more than the attracting behaviour, and the $y=0$ invariant manifold will not be asymptotically attracting (in the sense of condition (iii) above). For $\bar{x}$ close to, but greater than $\bar{x}_{cr}$, the $y=0$ invariant manifold will be asymptotically attracting yet small y-perturbations to typical trajectories in $y=0$ will cause the state vector to suffer a significant fraction of time in regions of phase space where the potential is repelling (in the sense of condition (iv) above).

The notion of riddled-basin dynamics is central to the parametrized model of quantum

spin-1/2 measurement put forward in section 3, where the attractors will correspond to observable states ("spin up / spin down"). However, the Sommerer and Ott example, is not quite appropriate for this purpose because the basin of attraction of the $y=\infty$ attractor is not itself everywhere riddled (for large $y$ the basin of attraction of the $y=\infty$ attractor has open sets) and therefore does not have the same character as the basin of the $y=0$ attractor. Rather, we require two symmetric chaotic attractors with mutually riddled or "intertwined" basins of attraction. Specifically, consider the potential

$$V(x,y) = V_Q(x,y) = (1-x^2)^2 + \varepsilon^2(x+\bar{x})\sin^2(\frac{y}{\varepsilon}) \qquad (2.5)$$

and substitute $V_Q(x,y)$ in place of $V_{SO}(x,y)$ in (2.1) with $\varepsilon \ll 1$. (In section 3 we will make a specific choice for $\varepsilon$ in terms of the parameters $\bar{x}$ and $x_{cr}$).

When $y/\varepsilon$ equals 0 and $\pi$, (2.1) and (2.5) reduce to the Duffing equations (2.3) and (2.4). Let us refer to the two chaotic attractors in these two invariant manifolds as $C_+$ and $C_-$ respectively. For $y/\varepsilon$ sufficiently close to either 0 or $\pi$, $V_Q$ is quadratic in $y$ and $V_Q(x,y) \approx V_{SO}(x,y)$. Hence, by earlier arguments, both $C_+$ and $C_-$ are attracting. Fom the symmetry properties of the potential, if the basin of attraction $b(C_+)$ is riddled, then so is $b(C_-)$. At $y/\varepsilon = \pi/2$ there is a third invariant manifold containing a chaotic attractor, $C_{\pi/2}$, equal, to $O(\varepsilon^2)$, to the Duffing attractor. However, using the arguments above, just as $C_+$, $C_-$ are attracting for points close to the invariant manifolds at $y/\varepsilon = 0$ and $y/\varepsilon = \pi$, $C_{\pi/2}$ will be repelling for points close to the $y/\varepsilon = \pi/2$ invariant manifold. Hereafter, we shall make the rescaling $y \mapsto y/\varepsilon$.

We illustrate the properties of the dynamical system (2.2), (2.6), based on numerical integrations using the Runge-Kutta algorithm. Specifically, we take as initial conditions a point $(x,y)$ at rest and integrate (2.1), (2.5) until it is virtually certain that $y$ has evolved to either $y=0$ and $y=\pi$. One parameter ("$TOL$") in the numerical algorithm, related to the timestep, is designed to control the accuracy of the calculation; under normal circumstances, a reduction in *TOL* leads to an approximately proportional reduction in the error of the solution. Let $S^{(n)}(x,y)=1$ if $y$ evolves to $0$, $S^{(n)}(x,y)=-1$ if $y$ evolves to $\pi$. Here $n=TOL_0/TOL$, where $TOL_0$ might be some reasonable a priori choice for *TOL*. Finally, let $S(x,y)=\pm 1$ if $y$ evolves to $0$ or $\pi$, respectively, according to an exact integration of (2.1) (2.5). (The meaning and indeed existence of an "exact integration" of (2.1), (2.5) is the focus of discussion in section 4). Note, by the symmetry of $V_Q$,

$$S(x,y)=-S(x,\pi+y) \qquad (2.6).$$

Fig 2a shows values of $S^{(1)}(x,y)$ for a regular grid of 200×200 points between $-1<x<1$, $0<y<\pi/2$. Here $\bar{x}=1.81$, other parameters have the values given above (with $\varepsilon=0.01$). Points shaded black have a value $S^{(1)}(x,y)=1$, unshaded points correspond to $S^{(1)}(x,y)=-1$. It can be seen that there is no particular pattern to the shading, except for a maximum of shaded points at $y=0$. The values $S^{(1)}(x,y)$ between $-1<x<1$, $\pi/2<y<2\pi$ can be obtained (in principle!) from Fig 2a and the symmetry relations (2.6). Fig 2b shows a similar calculation for a regular but finer grid of 200×200 points between $-.1<x<.1$, $.40\pi<y<.41\pi$. The calculation reveals finer (but still apparently random) detail for $S^{(1)}(x,y)$, not seen on the coarse grid. Complementary to this, Fig 2c shows a calculation of $S^{(n)}(x,.4\pi)$ for 200 regularly spaced points between $-1<x<1$ as $n$ increases from 1 to 200. The structure looks as random as that shown in Fig 2b. For any particular value $x$ there is no convergence in the value of $S^{(n)}(x,.4\pi)$ as the notional accuracy of the calculation increases. Moreover, there is no apparent correlation between values between different rows, eg if

$$C^{(n)(m)}=-\frac{1}{2}\int_{1}^{-1}S^{(n)}(x,.4\pi)S^{(m)}(x,.4\pi)dx \qquad (2.7)$$

then $C^{(n)(m)}\neq -1$, even though by construction $C=-\frac{1}{2}\int_{1}^{-1}S(x,.4\pi)S(x,.4\pi)dx=-1$. (The "-" sign is included in (2.7) in reference to the quantum correlation function (5.2).)

Fig 2 illustrates a fractal-like structure to the intertwined basins. As $\bar{x}$ approaches $x_{cr}$ from above, this apparently random structure persists at increasingly small scales, and the exact value $S(x,y)$ becomes increasingly difficult to compute. To emphasise this point further, we quote from Sommerer and Ott (1993): "although the underlying equations of motion are strictly deterministic, the riddled geometry of its basin structure, coupled with unavoidable perturbations, renders it effectively nondeterministic, and in the worst possible way". Such "unavoidable perturbations" could be thought of as numerical truncations errors. These authors conclude: "thus even qualitative reproducibility in simple classical systems cannot be taken for granted". Since we shall be attempting to apply this nonlinear property to describe physical reality, let us make the following pragmatic definition. If $\bar{x}$ is sufficiently close to $x_{cr}$ that it is impossible to determine with accuracy the basin of attraction of an arbitrary point in phase space from any physically-based algorithmic computation (ie within time or energy-momentum constraints imposed by the physical universe), then $S(x,y)$ can be said to be physically uncomputable.

Let us now define the functions $S_{\pm}(x,y)$, such that $S_{\pm}(x,y)=1$ when $S(x,y)=\pm 1$, respectively, and $S_{\pm}(x,y)=0$ otherwise (so $S(x,y)=S_{+}(x,y)-S_{-}(x,y)$). Then, even though $S(x,y)$ may be physically uncomputable, the probabilities $\int S_{\pm}(x,y)\rho_x(x)dx$ that $S(x,y)=\pm 1$, are themselves computable. Here $\rho_x(x)$ is a probability density function (PDF) of the state vector associated with the Duffing attractor. Specifically, $\rho_x(x)dx$ is the probability that an arbitrarily chosen point in the Duffing attractor on the line $dx/dt=0$ (see Fig 1), lies in the line element $dx$ at the point $x \in \mathbb{R}$. The line $dx/dt=0$ is used in this definition since the points $(x,y)$ are integrated from rest in calculating $S(x,y)$. For example, the fraction of unshaded points as a function of $y$ is shown in Fig 3 on a log-log scale for 400x400 points. Subject to fluctuations associated with sampling error, the calculation suggests that $\log[\int S_{\pm}(x,y)\rho_x(x)dx]$ converges to an approximately linear dependence on $\log y$.

This log-linear relationship is characteristic of riddled dynamics for small $\bar{x}-\bar{x}_{cr}$, at least for $y$-values sufficiently close to one of the invariant manifolds. (Ott el al, 1994). Let $\sigma^2$ be the variance of an ensemble of finite-time ($\Delta t$) estimates of a dominant transverse Lyapunov exponent $h_\perp$, taken over a large ensemble of random initial points near one of the invariant manifolds $M$. It can be shown that for sufficiently large $\Delta t$, $\sigma^2$ scales inversely with $\Delta t$. If we define $D$ as $\sigma^2 = 2\dfrac{D}{\Delta t}$ then the probability that a point at $y=y_0$ is attracted to $y=0$ scales as $(y_0)^\eta$ where $\eta = |h_\perp|/D$. Ott et al (1994) show that the evolution of an ensemble of points towards one of the attractors can be modelled as a diffusive random walk, and $D$ can be interpreted as a coefficient of diffusivity.

From Fig 3, $\eta \sim 0.2$, and if we write

$$\int S_+(x,y)\rho_x(x)dx = L(y) \qquad \int S_-(x,y)\rho_x(x)dx = 1-L(y) \qquad (2.8)$$

then to good accuracy

$$L(y) = 1 - \frac{1}{2}(2y/\pi)^\eta \quad 0 \leq y \leq \frac{\pi}{2} \qquad L(y) = \frac{1}{2}\left(\frac{2(\pi-y)}{\pi}\right)^\eta \quad \frac{\pi}{2} \leq y \leq \pi \qquad (2.9)$$

It should be noted that the functional dependence of $L$ on $y$ throughout the range $[0,\pi]$ does depend on the chosen parameters. For other values of $\bar{x}$, $\log L$ and $\log y$ are not linearly related throughout this range.

Equations (2.1), (2.5) will form the basis of the parametrized model of quantum spin measurement. However, $L(y)$ does not match the observed quantum spin probability function $\cos^2(y/2)$. This will be considered more carefully in a future paper taking $\bar{x}$ as a dynamical variable. However, for the purposes of this paper, it is sufficient to let $y'$ ($0 \leq y' \leq \pi$) be given by

$$L(y') = \cos^2(y/2) \qquad (2.10)$$

and define

$$Sp(x,y)=S(x,y') \qquad Sp_{\pm}(x,y)=S_{\pm}(x,y') \qquad (2.11)$$

Then from (2.8), (2.10) and (2.11)

$$\int Sp_{+}(x,y)\rho_{x}(x)dx=\cos^2(y/2) \qquad \int Sp_{-}(x,y)\rho_{x}(x)dx=\sin^2(y/2) \qquad (2.12)$$

Note that, from (2.6), (2.11), we have the symmetry relations

$$Sp(x,y)=-Sp(x,\pi+y) \qquad (2.13)$$

In Fig 2a we illustrated the function $S^{(1)}(x,y)$ for a regular grid of 200 x 200 regularly spaced points in $-1<x<1$, $0<y<\pi/2$. In Fig 4 we show the corresponding function $Sp^{(1)}(x,y)$, with $L(y)$ defined by (2.9), for the same regular grid of 200 x 200 points. The monotonic decrease of shaded points between $y=0$ and $y=\pi/2$ (consistent with (2.12)) can clearly be seen.

Now consider an ensemble of points with a total PDF of $\rho_{x}(x)\rho_{y}(y)$. The partial PDF $\rho_{x}(x)$ is as before. Let us consider two choices for $\rho_{y}(y)$. The first is where the points are chosen uniformly in $0 \leq y \leq 2\pi$ so that $\rho_{y}(y)=1/2\pi$ and

$$\iint Sp_{+}(x,y)\rho_{x}(x)\rho_{y}(y)dxdy = \frac{1}{2\pi}\iint Sp_{+}(x,y)\rho_{x}(x)dxdy$$
$$= \frac{1}{2\pi}\int \cos^2(y/2)dy = \frac{1}{2} \qquad (2.14)$$

The second is where all points are chosen on the line $y=y_0$ so that $\rho_{y}(y)=\delta(y-y_0)$ and

$$\iint Sp_{+}(x,y)\rho_{x}(x)\rho_{y}(y)dxdy = \int Sp_{+}(x,y_0)\rho_{x}(x)dx = \cos^2(y_0/2). \qquad (2.15)$$

In terms of the phase-space coordinates $X=(x,dx/dt)$, $Y=(y,dx/dt)$, the initial conditions for the calculations in Figs 2-4 were taken from the resting states $X=(x,0)$, $Y=(y,0)$. However, all results quoted so far are equally applicable if the integrations started at any point $X$ on the Duffing attractor. We can generalise the functions $S(X,y)$, $S_{\pm}(X,y)$ to have the same meanings as before, but to include arbitrary phase-space starting conditions $X$ on the Duffing attractor. The PDF $\rho_X(X)$ is now defined such that $\rho_X(X)dX$ is the probability that an arbitrary point on the Duffing attractor lies in a small area $dX$ centred on $X \in \mathbb{R}^2$.

### 3. A parametrized model for quantum spin-1/2 measurement

We now have the components to outline a parametrized model, "$Q$" as it shall be called, of quantum spin-1/2 measurement, motivated by the possible role and properties of gravitation, as outlined above. The qualification "parametrized" is in recognition of the fact that a complete theory of quantum gravitation has yet to be formulated. Moreover, there is no claim for the uniqueness or completeness of the equations for "$Q$" whose essential features are:

    i) determinism (ie no stochastically-defined terms)
    ii) a local-variable description of the state vector
    iii) time asymmetry
    iv) physical uncomputability

In section 5 we shall discuss Bohm's version of the EPR experiments. By way of preparation, in this section we apply $Q$ to the measurement of an ensemble of spin 1/2 particles using an Stern-Gerlach (SG) device with output directed to particles detectors "+" and "-". We presume these particles are all travelling along some given direction (the "$x$" axis).

In $Q$, the state vector parametrizes a spacelike 3-manifold containing both the single quantum particle, and the measuring apparatus. The result of a spin measurement on a given particle will be *determined* by the coordinates of the state vector in $\mathbb{R}^2 \times S^1$, together with the orientation of the SG device (and particle detectors) which perform the measurement. (The sense in which the word "determined" is meaningful is discussed in section 4.) Here $\mathbb{R}^2$ contains a chaotic attractor presumed to parametrize internal degrees of freedom associated with a "quantum gravitational" semi-group (gauge-group structure being ruled out by the Weyl Curvature Hypothesis). The "$S^1$" component describes a rotational degree of freedom in the (spatial) plane orthogonal to the $x$ axis relating to a local space-time isometry. As discussed below, the $\mathbb{R}^2$ and $S^1$ degrees of freedom become dynamically coupled during measurement.

We arbitrarily define a direction, the "$z$" axis, orthogonal to the $x$ axis, and represent a point in $\mathbb{R}^2 \times S^1$ by the pair $(\lambda,\mu)$ where $0 \leq \mu \leq 2\pi$ is an angular coordinate relative to the $z$ axis. Let us represent the probability of finding the state vector in a small volume $d\lambda d\mu$ at the point $(\lambda,\mu)$ in $\mathbb{R}^2 \times S^1$ by $\rho_\lambda(\lambda)\rho_\mu(\mu)d\lambda d\mu$. The partial PDF $\rho_\lambda(\lambda)$ will be determined solely by the (Cantor-set) geometry of the chaotic attractor. On the other hand, the partial PDF $\rho_\mu(\mu)$ will be determined by experimental design, eg whether the particle stream has been prepared by an SG device or not (defined more specifically below).

We now consider the process of measurement by an SG device oriented at an angle $\theta$ to the $z$ axis. We assume that this is associated with the proposed gravitational coupling between particle and measuring device as discussed in section 2c and references therein. In $Q$, this process will be defined in terms of a deterministic but uncomputable mapping $M_\theta$ taking $(\lambda,\mu)$ to either $(\lambda',\theta)$ or to $(\lambda',\pi+\theta)$. We define a function $Sp_\theta$ such that if $(\lambda,\mu)$ is mapped to $(\lambda',\theta)$ then $Sp_\theta(\lambda,\mu)=1$ ("spin up"), whilst if $(\lambda,\mu)$ is mapped to $(\lambda',\pi+\theta)$ then $Sp_\theta(\lambda,\mu)=-1$ ("spin down"). Obviously, we presume these spin functions to be normalised by $\hbar/2$. In the mapping $M_\theta$, $\lambda'$ is also a point on the Duffing attractor uncomputably related to $\lambda$, so the partial PDF $\rho_\lambda(\lambda)$ is unchanged by measurement. On the other hand, whatever the partial PDF $\rho_\mu(\mu)$ before measurement, it will comprise a linear combination of $\delta(\mu-\theta)$ and $\delta(\mu-\theta-\pi)$ after measurement.

We shall use the notion of intertwined attractor basins to make this mapping more explicit. As discussed in the previous section, such a model satisfies the properties set out at the beginning of this section. Specifically, the mapping $M_\theta$ is determined by (2.1) where $V=V_Q$ is given by (2.5) and we put $\varepsilon = \overline{x} - x_{cr}$. The $(\lambda,\mu)$ can be related to $(X,y)$ by

$$\lambda = X \qquad \mu = \theta + 2\cos^{-2} L(y) \qquad (3.1)$$

where $L(y)$ is the computable function defined in (2.8) and approximated by (2.9). (We note in passing that a more general relationship $\lambda = f(X,y)$ can also be defined.)

The topological undecidability of simplicial 4-manifolds is parametrized by the physical uncomputability associated with (2.1). The first term on the right hand side of (2.1a) and (2.1b) parametrizes the time asymmetry of quantum gravity by the posited Weyl Curvature Hypothesis. The second term on the right hand side of (2.1a) and (2.1b) parametrizes the essential nonlinearity of gravity. The final term on the right hand side of (2.1a) can be thought to determine an intrinsic time scale $2\pi/\omega$ associated with the quantum gravitational semi-group. It would seem plausible that $2\pi/\omega$ may be related to the Planck time. On this basis, (2.1) can be written by the discrete mapping

$$z_{n+1} = F(z_n) \qquad (3.2)$$

in the Planck-time related Poincaré surface of section $\omega\tau \bmod 2\pi = 0$.

The notion of an energy-dependent gravitationally-induced state reduction (as discussed in section 2c) can be parametrized in $Q$ by $\varepsilon = \bar{x} - x_{cr}$, which controls the value of the Lyapunov exponent transverse to the invariant manifolds. From (2.5), the $x$-gradient of $V_Q$ is independent of $y$ when $\bar{x} - x_{cr} = 0$, and the $x$-equation in (2.1) decouples from the $y$-variable, and reduces to the Duffing equation (2.3), (2.4). In terms of the unscaled $y$-variable in (2.5), (2.1) reduces to a linear equation with stationary solutions. Hence, modulo a quantum phase-factor, the $y$-equation in the (singular) limit $\bar{x} - x_{cr} = 0$ can be taken to represent pre-measurement Schrödinger dynamics.

For $\bar{x} - x_{cr} > 0$ the $\mathbb{R}^2$ and $S^1$ spaces become coupled and uncomputable state-vector evolution to one of the two possible attractors $C_+$, $C_-$ takes place. In a future paper we will consider more carefully the behaviour of the intertwined basin model as $\varepsilon$ is increased from

0. For the purposes of this paper, we imagine $\varepsilon$ to be fixed at a sufficiently small value that $S(x,y)$ is physically uncomputable. The timescale for evolution to one of the invariant manifolds is dependent on the size of $\varepsilon$.

For simplicity, we use the same values of the parameters as in section 2 (in turn based on values used by Sommerer and Ott, 1993). Before measurement, the spin state is presumed to be determined by $(\lambda,\mu)$. During measurement, the system evolves to either $\mu=\theta$ or $\mu=\pi+\theta$ mimicking (according to the results of Ott el al, 1994,) a random walk. The ensemble behaviour of this deterministic model will therefore appear (for all practical purposes) as a stochastic diffusive system. As mentioned in the introduction, there has been much work recently on interpreting quantum state vector reduction in terms of diffusive stochastic processes. In this respect, our results suggest that there may be a more fundamental deterministic theory underlying these stochastic models.

As in (2.11), see also the last paragraph of section 2d, we define the "spin function"
$$Sp_\theta(\lambda,\mu)=S(X,y) \tag{3.3}$$
and the partial functions
$$Sp_{\theta\pm}(\lambda,\mu)=S_\pm(X,y). \tag{3.4}$$
From (3.1) and (2.11), we can write
$$Sp_\theta(\lambda,\mu)=Sp_0(\lambda,\mu-\theta)\equiv Sp(\lambda,\mu-\theta). \tag{3.5}$$
From (2.13), we note the symmetries
$$Sp_\theta(\lambda,\mu)=-Sp_\theta(\lambda,\pi+\mu). \tag{3.6}$$

We now consider an ensemble of particles $(\lambda_i,\mu_i)$  $1\leq i\leq N$, and define the probability
$$Pr_+=\int\int Sp_{\theta+}(\lambda,\mu)\rho_\lambda(\lambda)\rho_\mu(\mu)d\lambda d\mu \tag{3.7}$$
From (2.8), (3.1) and (3.4) we have
$$\begin{aligned}Pr_+&=\int\int S_+(X,y)\rho_X(X)\rho_y(y)dXdy\\&=\int L(y)\rho_y(y)dy=\int\cos^2(\frac{\mu-\theta}{2})\rho_\mu(\mu)d\mu\end{aligned} \tag{3.8}$$

As in section 2d, let us consider two (experimentally-defined) choices for $\rho_\mu(\mu)$. Remember, we are considering a measurement of spin-1/2 particles by an SG device oriented at an angle $\theta$ to the $z$-axis. The first (isotropic) choice $\rho_\mu(\mu)=1/2\pi$ corresponds to the situation in which the stream of particles $(\lambda_i,\mu_i)$ has not been prepared in any way (eg by passing them through an SG device at an earlier time). In this situation

$$Pr_+ = \frac{1}{2\pi}\int \cos^2(\frac{\mu-\theta}{2})d\mu = \frac{1}{2}. \qquad (3.9)$$

For the second choice we consider a stream of particles emitted from one output channel of an SG device which is oriented at an angle $\phi$ with respect to the $z$ axis. For this particle stream, $\rho_\mu(\mu)=\delta(\mu-\phi)$, and

$$Pr_+ = \int \cos^2(\frac{\mu-\theta}{2})\delta(\mu-\phi)d\mu = \cos^2(\frac{\phi-\theta}{2}) \qquad (3.10)$$

It can be remarked that (3.10) can also be derived from (3.6) in the conventional quantum interpretation where the PDF $\rho_\mu(\mu)$ is the linear superposition of outcomes

$$\rho_\mu(\mu)=\cos^2(\frac{\phi-\theta}{2})\delta(\mu-\theta)+\sin^2(\frac{\phi-\theta}{2})\delta(\mu-\pi-\theta) \qquad (3.11)$$

Let us return to the $\lambda$ values associated with our ensemble of particles. By construction, $\lambda$ is a vector phase-space coordinate of the Duffing attractor. Under measurement, a point $\lambda$ on the Duffing attractor is mapped irreversibly and uncomputably by $M_\theta$ to a new point $\lambda'$ on the Duffing attractor. In Q, we consider that a correlated particle pair emitted from a zero angular momentum source can be treated as having the same pre-measurement $\lambda$ value. On the other hand, two uncorrelated particles will be treated as having different pre-measurement $\lambda$ values. Given a finite universe of spin 1/2 particles, we can presume that the chance that one particle's pre-measurement $\lambda$ value is *exactly* equal to the post-measurement $\lambda'$ associated with an earlier measurement on a different particle, is vanishingly small. Hence, if we consider three distinct spin-1/2 particles, at least one will be presumed to have a λ value different from the other two. This last sentence is fundamental to all that follows.

## 4. Truth Values and Natural Philosophy

Let us return to the issue mentioned briefly in section 2: the notion of existence of an exact integration of (2.1). Suppose we know the starting conditions $\lambda,\mu$ exactly. How can we determine in which basin of attraction $\lambda,\mu$ lies? First of all, a non-algorithmic mathematical method for integrating (2.1) from arbitrary initial conditions is neither known nor suspected. We can try to integrate (2.1) on a digital computer. However, since $Sp_\theta(\lambda,\mu)$ is physically uncomputable, the outcome $Sp_\theta^{(n)}(\lambda,\mu)$ is sensitive to numerical truncation errors, hence the sequence $\{Sp_\theta^{(n)}(\lambda,\mu)\}$ $n=1,2,...$, obtained from the computer will alternate irregularly between the values 1 and -1.

On the other hand, we have posited (2.1) as a component of physical theory. From the discussion in section 3, we assert that an exact integration of (2.1) is associated with a physical spin measurement $Sp_\theta(\lambda,\mu)$ on a spin-1/2 particle uniquely specified by $\lambda,\mu$; using a measuring device with specified orientation $\theta$. By performing this measurement, we are, in effect, asking "nature" to integrate (2.1). Such a measurement is, therefore, an exact "natural integration" of (2.1), (this phrase being used to constrast with the digital computer's inexact algorithmic integration). Since this type of measurement will lead to a definite result (spin up or spin down), we must clearly admit the existence of an exact non-algorithmic integration of (2.1), else $Q$ would be inconsistent with experiment at the most basic level. This natural integration therefore provides a value, $\pm 1$ for $Sp_{\theta_1}(\lambda,\mu)$. Hence the proposition $P_1: "S_{\theta_1}(\lambda,\mu)=1"$ must be definitely true or definitely false.

Now there is one fundamental difference between our posited natural integration of (2.1), and the algorithmic integrations illustrated in Fig. 2. From section 3, measurement involves an irreversible mapping $M_\theta:\lambda \to \lambda' \neq \lambda$. Hence, once a natural integration has been performed on a given particle with given $\lambda,\mu$ and apparatus orientation $\theta_1$, it is by definition impossible to perform (on that particle) a second natural integration with the same initial $\lambda,\mu$, and different orientation $\theta_2$. This contrasts with algorithmic integrations of (2.1) which (given sufficient computational resources) can be repeated for any given $\lambda,\mu$ with arbitrarily many different apparatus orientations $\theta_i$, $i=1,2....$

With this in mind, suppose we now ask the question: do there exist exact integrations of (2.1) from given $\lambda,\mu$, but correponding to two measurement orientations $\theta_1$, $\theta_2$. In other words, can the proposition $P_2$:"$S_{\theta_1}(\lambda,\mu)=1 \wedge S_{\theta_2}(\lambda,\mu)=1$" be said to be definitely true or definitely false? If sequences of physically-based algorithmic integrations do not determine a convergent solution for one orientation $\theta_1$, they will neither be able to determine pairs of convergent solutions for two orientations $\theta_1$, $\theta_2$. Moreover, from a physical point of view, we have just argued that it is not possible to determine the truth or falsehood of $P_2$ from measurements on a single particle. However, if a zero angular momentum source is used to produce correlated particle pairs, then, from section 3, a given pair can be described by the values $\lambda,\mu$ and $\lambda,\pi+\mu$. By measuring the spin of one particle with an SG apparatus aligned with orientation $\theta_1$, and by measuring the spin of the second particle with an SG apparatus aligned with orientation $\theta_2$, and by using (3.6), we can determine the truth or falsehood of $P_2$ by natural integration. Hence $P_2$ can also be said to be definitely true, or definitely false.

Let us consider one last, but (for sections 5 and 6) critical question. For given $\lambda,\mu$, do there exist triples of exact integrations of (2.1) for three different orientations $\theta_1$, $\theta_2$, $\theta_3$? In other words, can the proposition $P_3$:"$S_{\theta_1}(\lambda,\mu)=1 \wedge S_{\theta_2}(\lambda,\mu)=1 \wedge S_{\theta_3}(\lambda,\mu)=1$" be said to be definitely true or definitely false? Once we ask for this third solution (for given $\lambda,\mu$) then we are no longer able to perform an exact natural integration from either single isolated particles, or from correlated particle pairs. From the discussion in section 3, there is no physical experiment that can determine the triples of values that are required to validate or falsify $P_3$. In other words, $P_3$ is a well-defined proposition within the formalism of $Q$, but has no place in the physical universe described by $Q$. Since such triples of integrations cannot be shown to exist physically, computationally or mathematically, then one may question whether, within $Q$, it is meaningful to say that $P_3$ has a definite truth value.

From a philosophical point of view we are attempting here to define an approach to the metaphysical notion of existence, for circumstances in which components of physical theory are neither physically computable nor mathematically deducible. There is a parallel here with the debate about the notion of mathematical existence (still an active area, see eg Henle, 1991; Lambeck, 1994). Consider, for example, the Mathematical Formalist (hereafter MF) approach based on a description of mathematics as a formal language, a collection of axioms, and a means of deduction (Benacerraf and Putnam, 1984). MF truth is identified with those theorems which follow from the axioms in a finite number of steps, and is therefore entirely dependent on the chosen language, axioms and rules of inference. For example, in a standard formal system (such as Zermelo-Frankael set theory), the Continuum Hypothesis is regarded as neither true nor false, since either it or its negation may be added as axioms without inconsistency. The correspondence between such formally-undecidable propositions and time-integrals in uncomputable dynamical systems theory, may be quite close (Moore, 1990).

What is proposed here is an adaptation of the MF approach. In this adaptation, a proposition from physical theory is treated as "true" if it can be proved either by mathematical deductive techniques, or by numerical integration using a physically-based algorithmic solver, or verified by natural integration (physical measurement). Conversely, that proposition is "false" if it can be disproved mathematically, by numerical integration or falsified experimentally. However, a proposition which can be proved neither by mathematical deduction, nor by algorithmic integration, nor by natural integration, would be treated as having no well-defined truth value. Obviously, consistent with scientific methodology, those (decidable) propositions which can be tested by conventional deductive technniques, must be consistent with results from natural integrations. (The probabilities (3.10) of measuring spin up or spin down are examples of the latter.)

MF is a conventional approach to existence in Mathematical Philosophy; we shall refer to this adaptation to Natural Philosophy as "Natural Formalism" (hereafter NF). In particular, the existence of a definite truth value for $P_i$ $i \geq 3$ is denied in NF.

Now suppose we perform the spin measurements $Sp_{\theta_1}(\lambda,\mu)$ and $Sp_{\theta_2}(\lambda,\mu)$ on a given correlated particle pair. Having performed this experiment, is it valid to make the following statement: Bearing in mind the results $Sp_{\theta_1}(\lambda,\mu)$ and $Sp_{\theta_2}(\lambda,\mu)$ already obtained, if we had performed a different measurement $Sp_{\theta_3}(\lambda,\mu)$ on one of the particles, then the result would definitely have been either "up" or "down"? Whilst this statement may seem at first sight to be rather uncontentious, it presupposes the existence of a definite (albeit unknown) truth value for $P_3$ (a definite value for $P_2$ having previously been obtained).

The purpose of this is to show (within an NF interpretation of $Q$) the possible invalidity of counterfactual statements (Lewis, 1976) such as: "If measurement $M$ were to be performed and its outcome were to be $O_1$, then if, instead of $M$, $N$ were to be performed, its outcome would be $O_2$". As discussed in the next section, counterfactual validity has been used as the basis of a proof of Bell's theorem (Stapp, 1994) in which no explicit reference to hidden-variable theories is made. In these references, the use of counterfactual statements is related to a more primitive notion known as "closeness of possible worlds". By contrast, within an NF interpretation of $Q$, a natural integration of (2.1) from given $\lambda,\mu$ is unique and irreversible. Once performed, then additional or hypothetical alternative integral solutions from the same $\lambda,\mu$ do not exist. Hence the hypothetical worlds in which these hypothetical alternative measurements take place also do not exist.

There are are some intriguing consequences of the NF view. Firstly NF appears to embrace Gödel's theorem in a way in which MF cannot. Specifically, MF is often criticised (eg Penrose, 1989) since it does not appear consistent with Gödel's theorem (which states that there are mathematical propositions which are true but cannot be proved to be true). However, there is no inconsistency between NF and Gödel's theorem. For example, the proposition $P_1$ cannot be proved by mathematical analysis or algorithmic integration, though can be verified or falsified by natural integration. Hence, in NF, either $P_1$ or its negation is true, though neither can be proved to be true by conventional deductive logic. In more emotive language, Gödel's theorem appears to proclaim the existence of an uncomputable physical universe!

A second remark concerns the relationship with a Mathematical Platonist point of view, in which a primal notion is that truth is not internal, but rather is independent of the formal system under study. In more emotive language, Platonist truth is deemed to be "out there somewhere" (a quotation from Davis and Hersh, 1990). However, if $P_3$ is not part of the physical universe, it is, by definition, not "out there somewhere". Denying the truth or falsehood of $P_3$ may, therefore, not be contrary to an attempt to apply Platonist thinking to possible noncomputable physical theory.

Finally, let us make some remarks about the relationship between NF and more familiar quantum interpretations (eg Rae, 1992). Firstly, in view of the discussion above, the uniqueness of physical reality in the NF approach to $Q$ is antithetical to the many-worlds interpretation of quantum theory. Secondly, although the primary motivation of this paper is to support Einstein's view that physics is local and deterministic, NF has some elements in common with the standard Copenhagen interpretation. For example, in both approaches it is meaningless to ask for the outcome of an unperformable experiment. On the other hand, NF does not insist that a quantity be considered "real" only if it has been measured, or is in a situation where the outcome of an experiment is predictable. Just as the Continuum Hypothesis is a perfectly well-defined and hence "real" proposition in standard set theory, so $P_1$ is also a well-defined and "real" proposition in $Q$.

## 5. Bell's theorem

In the discussion in section 3, the physically uncomputable nature of $Q$ was not fundamental; the statistics of local quantum measurement could have been obtained by taking one of the computable spin functions $Sp_\theta^{(n)}(\lambda,\mu)$. However, such a model, when applied to the description of physically-separated quantum measurements, would necessarily have to satisfy Bell's inequalities, and therefore be incompatible with experimental data (Aspect and Grangier, 1986). In this section we discuss the standard derivation of Bell's inequalities using the uncomputable function $Sp_\theta(\lambda,\mu)$ to describe measurement, bearing in mind the discussion in the previous sections.

We consider an experiment in which pairs of spin *1/2* particles are produced from a zero angular momentum source. For the first particle stream, the *z* component of spin is measured with an SG device together with "+" and "-" counters. For the second particle stream, the spin at an angle $\phi$ to the *z* axis is similarly measured. Let us assume $Q$, and represent the first particle stream (before measurement) by the pairs $(\lambda_i,\mu_i)$ $i=1,2,...N$, and the second particle stream (also before measurement) by the pairs $(\lambda_i,\pi+\mu_i)$ $i=1,2,...N$. According to $Q$, after measurement, the spin value of the *i* th particle in the first stream is $Sp_0(\lambda_i,\mu_i)$, the spin value of the *i* th particle in the second stream is $Sp_\phi(\lambda_i,\pi+\mu_i)$. The correlation function

$$C(\phi) = \frac{1}{N}\sum_{i=1}^{i=N} Sp_0(\lambda_i,\mu_i) Sp_\phi(\lambda_i,\pi+\mu_i) \tag{5.1}$$

is determined from the measurements and becomes independent of $N$ for sufficiently large $N$. From (3.6), this can be written as

$$C(\phi) = -\frac{1}{N}\sum_{i=1}^{i=N} Sp_0(\lambda_i,\mu_i) Sp_\phi(\lambda_i,\mu_i) \tag{5.2}$$

In a standard derivation of Bell's theorem, (eg Rae, 1992), it is now assumed that *if*

*the second measuring apparatus had been set up with orientation* θ, *(having in fact been set up with orientation* ϕ*) and used to measure the N particles of the second particle stream, then the result would have been*

$$C(\theta) = -\frac{1}{N} \sum_{i=1}^{i=N} Sp_0(\lambda_i, \mu_i) Sp_\theta(\lambda_i, \mu_i) \tag{5.3}$$

so that

$$C(\phi) - C(\theta) = -\frac{1}{N} \sum_{i=1}^{i=N} Sp_0(\lambda_i, \mu_i) [Sp_\phi(\lambda_i, \mu_i) - Sp_\theta(\lambda_i, \mu_i)] \tag{5.4}$$

Alternatively, if we write (5.2) by the continuum equivalent,

$$C(\phi) = -\iint Sp_0(\lambda,\mu) Sp_\phi(\lambda,\mu) \rho_\lambda(\lambda) \rho_\mu(\mu) d\lambda d\mu \tag{5.5}$$

the assumption in italics above implies that we can also write

$$C(\theta) = -\iint Sp_0(\lambda,\mu) Sp_\theta(\lambda,\mu) \rho_\lambda(\lambda) \rho_\mu(\mu) d\lambda d\mu \tag{5.6}$$

so that

$$C(\phi) - C(\theta) = -\iint Sp_0(\lambda,\mu) [Sp_\phi(\lambda,\mu) - Sp_\theta(\lambda,\mu)] \rho_\lambda(\lambda) \rho_\mu(\mu) d\lambda d\mu, \tag{5.7}$$

From (5.4) or (5.7), it is easy to derive the Bell inequality

$$|C(\phi) - C(\theta)| - C(\theta - \phi) \leq 1 \tag{5.8}$$

Generalisations of (5.8) have been shown to be inconsistent with experiment (Aspect and Grangier, 1986), and from this it is usually inferred that the assumptions of locality and determinism are together necessarily false.

However, note that in order to evaluate either the summand of (5.4), or the integrand of (5.7), we require the triple of values $Sp_0(\lambda,\mu)$, $Sp_\phi(\lambda,\mu)$, $Sp_\theta(\lambda,\mu)$ for given $\lambda,\mu$. If $Sp_\theta(\lambda,\mu)$ were computable, then such triples unquestionably exist, and Bell's theorem (5.8) can be established. This assumption of unambiguous existence is implicit in proving the invalidity of conventional (Bohm-type) hidden-variable theories. However, according to the discussion in section 4, the mathematical existence of such triples, or equivalently the truth values of $P_3$, in uncomputable $Q$ (where a general non-algorithmic mathematical integration of (2.1) is neither known nor suspected) is ambiguous. Since, from the discussion in section 4, $P_3$ has no place in the physical universe described by $Q$, then according to the NF interpretation of $Q$, such triples are denied existence and Bell's theorem (5.8) cannot be established, locality and determinism notwithstanding.

Of course, $C(\phi)$, $C(\theta)$ and $C(\theta-\phi)$ are all individually well defined in NF from distinct particle ensembles. For example, if an experiment is performed on a second ensemble of $N$ particle pairs $N+1 \leq i \leq 2N$, with the first measuring apparatus aligned with the $z$-axis, the second oriented at the angle $\theta$ then both $C(\phi)$ given by (5.1) and

$$C(\theta) = -\frac{1}{N} \sum_{i=N+1}^{i=2N} Sp_0(\lambda_i,\mu_i) Sp_\theta(\lambda_i,\mu_i) \qquad (5.9)$$

are each well defined in NF.

As noted in section 4, a version of Bell's theorem has been proven which assumes neither hidden variables nor determinism, but assumes the validity of counterfactual statements (Stapp, 1994). For example, the italicised clause above is essentially counterfactual. However, as discussed in section 4, the validity of such counterfactual statements in an NF interpretation of $Q$ is denied, even though $Q$ is deterministic. As such, the reformulation of Bell's theorem in terms purely of counterfactuals does not necessarily prevent the arguments of this paper applying.

We have shown that $Q$'s locality and determinism does not necessarily lead to an inconsistency with Bell's theorem. However, we have not deduced mathematically from $Q$ the observed correlation $C(\theta) = -\cos\theta$. Certainly, from section 2, $C(\theta) = -\cos\theta$ is not deducible from a physically-based algorithm, since the correlations are not computable. However, it is possible that a non-algorithmic deduction of the ensemble correlations may exist. It is intended to investigate this more thoroughly in the future. However, as stated in the introduction to this paper, the purpose of the present study is not to put forward a complete quantum theory, but rather to challenge the conventional assumption that quantum physics is inconsistent with locality and determinism. The value of correlation integrals is not central to this debate. In order to emphasise further this aspect, we discuss in the next section a quantum entanglement theorem in which ensemble statistics are not directly relevant at all.

6. **Magic dodecahedra; a new example of quantum non-locality?**

Recently, Penrose (1994a,b) has discussed a "gedanken" device, entirely consistent in principle with quantum physics, with which to demonstrate a version of the Bell-Kochen-Specker entanglement paradox (Heywood and Redhead, 1983; Brown and Svetlichny, 1990) and hence apparent quantum nonlocality. Unlike the account of Bell's theorem above, no ensemble averaging enters the discussion, and in particular, the question of the value of the correlation function $C(\theta)$ is not directly relevant. In this section we analyse whether indeed this construction implies non-locality in an NF interpretation of $Q$.

The device is a dodecahedron, supposedly manufactured by some advanced civilisation. Two spin 3/2 atoms are prepared with an initial combined total spin of 0, carefully separated, isolated, and placed in the centre of two identical dodecahedra. These devices are then sent to distant colleagues, who orient their dodecahedra so that they are aligned perfectly (eg with respect to some distant star).

At each dodecahedron vertex is a button, which, if pressed, initiates a partial spin measurement which determines whether the $m$-value of the atomic spin in the direction out from the centre to that particular vertex is $+1/2$ (rather than one of the alternative possible values +3/2, -1/2 and -3/2). If the measured spin value is 1/2, the dodecahedron is irreversibly destroyed by some internal mechanism (and no further measurements are possible). Otherwise, the dodecahedron remains intact (a "null" measurement), and is available for further partial spin measurements.

The two colleagues are invited to independently select one vertex and press, in some arbitrary order, the buttons associated with each of the three vertices adjacent to the selected one. From the quantum mechanics of angular momentum, the $m$-value of the spin in the selected direction can be deduced from button pushes at the adjacent vertices. Moreover, the manufacturers of the dodecahedra can guarantee the following:

a)  If the two colleagues select diametrically opposite vertices, and if pressing one particular adjacent button destroys one colleague's dodecahedron, then the diametrically-opposite button on the other colleague's dodecahedron will similarly destroy that device.

b)  If the two colleagues happen to select exactly corresponding vertices, then at least one dodecahedron must be destroyed by one of the six possible button pushes by the two colleagues.

If we assume there are no long-distance influences relating their two dodecahedra, then, according to Penrose (1994a,b), the two colleagues can individually make the following deductions about their own dodecahedron:

c)  each of that dodecahedron's vertices must be pre-assigned as either null or destructive by the manufacturers,

d)  no two next-to-adjacent vertices can be both destructive,

e)  no set of six vertices adjacent to a pair of antipodal ones can all be null.

However, from d)-e), one can show that it is impossible to label each of the vertices of a dodecahedron as either "destructive" or "null". Hence there is an inconsistency with c). Penrose concludes that the assumption of locality (no long-distance influence) must be incorrect.

Penrose deduces (c) from the fact that the two colleagues might happen to select diametrically-opposite vertices. The manufacturers of the dodecahedra cannot know this in advance. Thus, it is argued, if a particular button press by one colleague destroys his dodecahedron, then the manufacturers must have pre-arranged the other colleagues' vertex to be destructive in order to be consistent with (a).

Let us examine this argument further in an NF interpretation of $Q$. For such a system, we should have to extend $Q$ to take account of four attractors $C_i$ $1 \leq i \leq 4$, corresponding to the four possible spin values, whilst retaining locality, determinism and uncomputability. We shall not attempt a detailed description of this extension here. On the basis of results in section 2, the difficulties involved in formulating this extension precisely, appear to be solely technical.

As before, the basins of attraction of the $C_i$ must together cover phase space, and we presume that the basin of any one of these attractors is intertwined with respect to its complement. Let $\lambda,\mu$ determine the $m$-value of the spin state $Sp_{\theta_n}(\lambda,\mu)$ of one of the spin 3/2 atoms, and $\lambda,\pi+\mu$ determine the spin state $Sp_{\theta_n}(\lambda,\pi+\mu)$ of the other atom, where $\theta_n$ is the orientation (with respect to the distant star) of one of the possible dodecahedron vertices. The guarantee a) can be met if

$$Sp_{\theta_n}(\lambda,\mu) = -Sp_{\theta_n+\pi}(\lambda,\mu), \qquad (6.1)$$

$$Sp_{\theta_n}(\lambda,\mu) = -Sp_{\theta_n}(\lambda,\pi+\mu), \qquad (6.2)$$

Note that (6.1), (6.2) are closely related to the conditions (3.5), (3.6) for the spin-1/2 model.

In quantum mechanics, guarantee b) and deductions d) and e) are, in addition, associated with the orthogonality of spin states determined by button pushes of any two next-to-adjacent vertices $\theta_n$, $\theta_m$. These can be satisfied in "extended $Q$" if

$$Sp_{\theta_n}(\lambda,\mu) \neq Sp_{\theta_m}(\lambda,\mu). \qquad (6.3)$$

Now in extended $Q$, pressing a button on a vertex with orientation $\theta_n$ determines the truth or falsehood of the proposition $P_{\theta_n}:^{//} Sp_{\theta_n}(\lambda,\mu)=1/2^{//}$, through some natural integration of the dynamical equations of extended $Q$. If $P_{\theta_n}$ is true, the dodecahedron is destroyed (the natural integration is complete and irreversible).  However, since each basin and its complement are intertwined, the propositions $P_{\theta_n}$ are effectively undecidable, despite the functions $Sp_{\theta_n}(\lambda,\mu)$ being deterministic. As in $Q$, we assume that, for general $(\lambda,\mu)$, there are no non-algorithmic mathematical solutions to such propositions, so that the existence to definite truth values to $P_{\theta_n}$ is ambiguous for propositions which cannot be tested by any physical experiment. (The possibility of non-existence of truth values is not prejudiced by the relations (6.1)-(6.3). For example, from (6.1) if $P_{\theta_n}$ is true then $P_{\theta_n+\pi}$ is false. However, if $P_{\theta_n}$ has no definite truth value, then neither does $P_{\theta_n+\pi}$.) We wish to show that in attempting to demonstrate c), we are forced to consider such unprovable and untestable propositions.

Suppose one of the colleagues pre-selects two different vertices $S_1$ and $S_2$ at random, and then finally decides only to press buttons on vertices adjacent to $S_1$. Suppose, on one of the three possible button pushes, the dodecahedron is destroyed. This constitutes a natural integration of extended $Q$. From (6.3), the other two vertices adjacent to $S_1$ would have been null if they had been pushed. Now suppose $S_2$ had been adjacent to one of these null vertices. After the buttons had been pushed, would it be valid for the colleague to make the following statement: "If the button corresponding to $S_2$ were to have been pressed (the buttons adjacent to $S_1$ having in fact been pressed, and one of them found to be destructive), then the result would have either been null or destructive." ?

However unexceptionable this statement may appear at first sight,  it is a counterfactual statement of the type discussed in section 4. In section 4 we denied the validity of this type of statement in an NF interpretation of $Q$. In particular,  $P_\phi$, where $\phi$ denotes the orientation of $S_2$, is deemed to have no definite truth value in NF since (by supposition) it is neither amenable to any form of mathematical deduction, nor to validation by natural integration (the dodecahedron was destroyed by a button-push of a vertex adjacent to $S_1$). Hence, with specific reference to $S_2$, *it is not the case that each of the dodecahedron's* *vertices must have been pre-assigned as either null or destructive.*

This result is equivalent to the NF-based invalidity of $P_3$ in section 3. Each colleague can only perform one irreversible measurement on his dodecahedron; together they are allowed at most two. The possibility of three destructive measurements is not part of the physical universe inhabited by the two colleagues and their dodecahedra.

Let us consider one last possibility. Again suppose a colleague selects a vertex at random, and suppose all three buttons on adjacent vertices give null results. He now selects a new vertex and presses the corresponding adjacent buttons. Suppose again all are null. Could the colleague procede in this way, pressing only null buttons, leaving him able to deduce all the dangerous buttons from d) and e) without pressing them? The answer is no. If he were able to do this he would have found a complete labelling of the dodecahedron which we know to be impossible. Hence, we are assured (from d), e), and the dodecahedron geometry) that one of the button pushes must be destructive, before a complete labelling can be unambiguouslyng deduced. Therefore, by pressing buttons, ie performing natural integrations, and using deductive logic based on (6.1)-(6.3), one need not come into conflict with the assumption of locality.

In conclusion, the Bell-Kochen-Specker paradox (as revealed by Penrose's magic dodecahedron gedanken device), does not imply nonlocality within a NF interpretation of (extended) $Q$.

## 7. Conclusions

The principal purpose of this paper is to challenge the conventional "text-book" view, that Einstein was wrong to believe that quantum physics is local and deterministic. To do so, we have developed a model, $Q$, which is local and deterministic, can describe the observed statistics associated with the measurement of spin 1/2 particles, yet is not necessarily constrained by either Bell's theorem, or other quantum entanglement theorems. Appropriately, the fundamental physical process that may admit Einstein's view about the nature of the physical world is gravity.

However, we have made use of theory that would not have been available to Einstein. In particular, we have drawn on the existence of dissipative dynamical systems with multiple attractors whose basins are intertwined (they cover phase space, but have no open sets). Although deterministic (the equations of motion contain no stochastically-defined terms), for suitable parameter values, the evolution of the state vector of such systems is physically uncomputable. According to $Q$, God has no need for dice, though cosmos-bound mortals may as well continue to use them!

An additional concept that, certainly in Einstein's day, would not have been considered a relevant issue in the formulation of a physical theory is that of the existence of definite truth values associated with a well-posed proposition from that theory. Since conventional physical theories are computable, any well-posed proposition from such a theory can, in principle, be shown to be true or false. However, in $Q$, there are many propositions that cannot be shown to be either true or false by physically-based algorithmic calculation. Moreover, there are no grounds to believe that general non-algorithmic solutions to such propositions exist. Of these propositions, some may correspond to physically testable experiments, others may not. The quantum entanglement theorems considered in this paper involve (counterfactual) propositions of the latter type. The approach developed in this paper, in some ways an adaptation of mathematical formalism to physical theory, denies definite truth values to the latter type of well-posed proposition.

Possibly this approach provides some basis for reconciling Bohr and Einstein's views on the interpretation of quantum theory, at least in the sense that it is meaningless to ask for the truth of a undecidable proposition corresponding to an unperformable experiment. However, the approach developed here does not insist that a quantity be considered "real" only if it has been measured or is in a situation where the outcome of an experiment is completely predictable. Based on $Q$, we cannot yet rule out a "reality in space and time, free from the spooky action at a distance".

## Acknowledgements

I am very grateful to Dr H.R. Brown, Professor R.Penrose, and Professor I.N.Percival for taking the time to disucss with me the issues put forward in this paper. These discussions helped clarify my thinking and led to a more carefully-reasoned manuscript. An anonymous reviewer made helpful suggestions for improving the original manuscript.

Figure Captions

1. The Duffing twin well attractor in the phase space ($x$, $dx/dt$) obtained by integrating numerically the differential equation (2.3).

2. a) An illustration of $S^{(1)}(x,y)$ for 200 x 200 uniformly spaced points in the range $-1<x<1$, $0<y<\pi/2$. Black squares correspond to values $S^{(1)}(x,y)=1$, white squares to $S^{(1)}(x,y)=-1$. The integration is continued until it is virtually certain that a point in phase space has been captured by either $C_+$ or $C_-$. b) as a) but for 200 x 200 uniformly spaced points in the range $-0.1<x<0.1$, $.40\pi<y<.41\pi$. c) an illustration of $S^{(n)}(x,.4\pi)$ for 200 uniformly spaced points in the range $-1<x<1$, and for $n=1,2,...,200$.

3. The logarithm of the probability that $S^{(1)}(x,y)=1$, estimated from 400 uniformly spaced points in the range $-1<x<1$, that evolve to y=0, as a function of 400 points in the y direction, spaced uniformly in log $y$ between $0<y<\pi/2$.

4. An illustration of $Sp^{(1)}(x,y)$ for 200 x 200 uniformly spaced points in the range $-1<x<1$, $0<y<\pi/2$. Black squares correspond to values $Sp^{(1)}(x,y)=1$, white squares to $Sp^{(1)}(x,y)=-1$.